\documentclass{article}
\usepackage[T1]{fontenc} 
\usepackage[utf8]{inputenc} 
\usepackage{ismir,amsmath,cite,url}
\usepackage{graphicx}
\usepackage{color}
\usepackage{multirow}

\usepackage{subcaption}
\usepackage{amssymb}
\usepackage{multirow}
\usepackage{graphicx}
\usepackage{makecell}

\def\thline{\noalign{\hrule height 1.0pt}}

\title{Towards Musically Meaningful Explanations \\ Using Source Separation}

\multauthor
{Verena Haunschmid$^1$ \hspace{1cm} Ethan Manilow$^3$ \hspace{1cm} Gerhard Widmer$^{1,2}$}{
 $^1$ Institute of Computational Perception, Johannes Kepler University, Linz, Austria\\
$^2$ LIT Artificial Intelligence Lab, Johannes Kepler University, Linz, Austria\\
$^3$  Interactive Audio Lab, Northwestern University, Evanston, IL, USA\\
{\tt\small verena.haunschmid@jku.at}
}

\begin{document}

\maketitle

\begin{abstract}
Deep neural networks (DNNs) are successfully applied in a wide variety of music information retrieval (MIR) tasks. Such models are usually considered "black boxes", meaning that their predictions are not interpretable.
Prior work on explainable models in MIR has generally used image processing tools
to produce explanations for DNN predictions, but these are not necessarily musically meaningful,
or can be listened to (which, arguably, is important in music).
We propose \textit{audioLIME}, a method based on Local Interpretable Model-agnostic Explanation (LIME), extended by a musical definition of locality. LIME learns locally linear models on perturbations of an example that we want to explain. Instead of extracting components of the spectrogram using image segmentation as part of the LIME pipeline, we propose using source separation. The perturbations are created by switching on/off sources which makes our explanations listenable.
We first validate audioLIME on a classifier that was deliberately trained to confuse the true target with a spurious signal, and show that this can easily be detected using our method. We then show that it passes a sanity check that many available explanation methods fail. Finally, we demonstrate the general applicability of our (model-agnostic) method on a third-party music tagger.
\end{abstract}

\section{Introduction}

Deep neural networks are used in a wide variety of music information retrieval (MIR) tasks. While they generally achieve great results according to standard metrics, they  
are notoriously considered ``black box'' systems, meaning that it is hard to interpret how or why they determine their output. This can lead to problems where a network does not learn what its designers intend. For instance, a system that outperformed many other models on a popular genre classification dataset actually benefitted from information below 20 Hz, which should be irrelevant because such information is inaudible to humans~\cite{Rodriguez-Algarra16WhereIsTheMusic}. Prior to that, Sturm showed that MIR systems can perform well on a task without understanding anything about music~\cite{Sturm14Horses}.

As increasingly sophisticated machine learning systems become more pervasive in the life cycle of musical artifacts--from production tools, to distribution platforms, to discovery and recommendation engines--the stakeholders of these systems (such as musicians, sound engineers, machine learning practitioners, and whole corporations) must be able to trust these systems to produce reliable results. Research in interpretable machine learning aims to elucidate these models and build trust in these systems.

In recent years several tools from the field of interpretable machine learning have found their way into MIR. Most published works in this field use approaches from image processing and apply them to spectrograms~\cite{MishraSD17SoundLIME,Haunschmid19TwoLevel,MishraSD18FeatureInversion,MishraSD18Explaining}. Other works claim interpretability arises as a natural consequence of a particular network architecture, such as attention mechanisms~\cite{Won19SelfAttention,Won19VisualizingSelfAttention} or invertible neural networks~\cite{KelzW19Invertible}. However, whether attention mechanisms are explanations is currently critically debated in the machine learning community~\cite{JainW19Attention, WiegreffeP19Attention}. Regardless, while it may be preferable to explicitly design models in an interpretable way, models will always exist without any internal interpretability mechanisms, and thus there will always be a need for external explainability tools.

The goal of this work is to propose a new method that creates \textit{model-agnostic}, \textit{post-hoc} explanations for arbitrary classifiers/taggers in MIR, which should also be \textit{listenable} and, if possible, musically meaningful. To achieve this, we build on \textit{Local Interpretable Model-agnostic Explanations (LIME)}~\cite{Ribeiro0G16}, a popular tool for interpreting text and image classifiers, and propose a new notion of ``locality'', creating perturbations based on source separation estimates (rather than applying image segmentation to spectrograms, as in~\cite{Haunschmid19TwoLevel,MishraSD17SoundLIME}). We call our new method \textit{audioLIME}.
\footnote{\url{https://github.com/CPJKU/audioLIME}}

Experimental evaluation of the method will proceed in several stages. In the first stage, we will verify audioLIME with musical examples that are generated in a controlled way. This is intended to check whether our approach can be used to explain the predictions of a classifier that was deliberately trained on confounding data. We will see that the explanations of wrongly and (maybe surprisingly) correctly classified cases will indeed reveal that the model has learned to relate the confounding source with the prediction target.
Additionally, we will perform a simple sanity check~\cite{AdebayoGMGHK18SanityChecks} in order to show that our method does not deliver meaningful `explanations' when applied to a randomly initialized, untrained model. 
In the second stage, we will demonstrate the general applicability of our method on a publicly available music auto-tagger. 
In the final stage, we will show how we can also perform a \textit{global analysis} of a prediction model. We show this using a convenient property of our explanations: the fact that each example is explained by the same components (estimated sources). 


\section{Musically Meaningful Explanations with audioLIME}
\label{sec:audioLIME}

audioLIME is based on the general LIME~\cite{Ribeiro0G16} framework and extends its definition of locality for musical data. This section first gives an overview of the original definition of LIME and then defines locality in a musical sense.

\subsection{Local Interpretable Model-Agnostic Explanations (LIME)}
\label{subsec:LIME}

Using local surrogate models in order to explain the predictions of a global model was first introduced in~\cite{Ribeiro0G16}. LIME creates explanations for a test example $x \in \mathbb{R}^d$ by learning a simple model locally around that example. The first step in the LIME algorithm is the creation of an interpretable representation $x' \in \{0, 1\}^{d'}$ where $d'$ represents the number of interpretable \textit{components}. Components are features designed to be easily understood by humans, e.g. single words in text classification or superpixels (a contiguous patch of pixels) in image classification. 0 and 1 denote their absence or presence, respectively.
Using the interpretable representation, $n$ perturbations of the test example are created in the domain $\{0, 1\}^{d'}$ resulting in a potentially large number of binary vectors ($z_1', \ldots, z_n'$). These interpretable representations are mapped back to $\mathbb{R}^d$ (the input space)\footnote{For an image $z$, example $z'=\{0, 1, 0, 1\}$ results in an image where the first and third (of four) superpixels are replaced by gray patches.}. Those perturbed samples (e.g. images) are fed into the global model. The interpretable representations $(z_1', \ldots, z_n')$ and predictions of the global model $(f(z_1), \ldots, f(z_n))$ are used as features and labels to train a linear model, whose weights serve as an explanation for the prediction of the global model on the test example~\cite{Ribeiro0G16}.

\begin{figure}[!tb]
    \centering
    \includegraphics[width=0.9\columnwidth]{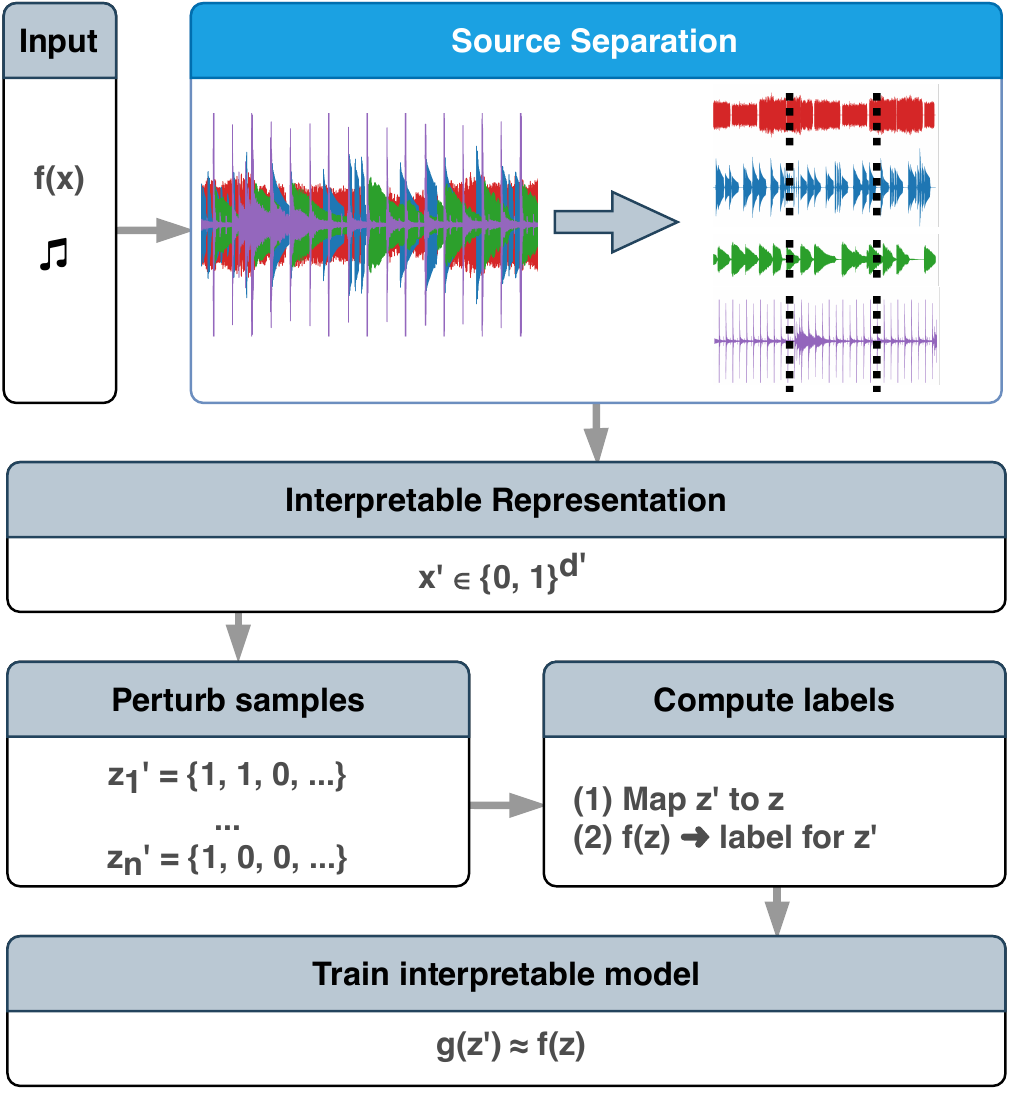}
    \caption{audioLIME closely follows the general LIME pipeline. The difference is shown in the blue box (``Source Separation''). The input audio is decomposed into $d' = C \times \tau$ components ($C$ sources, $\tau$ time segments).}
    \label{fig:audioLIME}
\end{figure}

LIME is also suited for classifiers/taggers in MIR, but the definition of interpretable components has to be adapted to this domain. Mishra et al.~\cite{MishraSD17SoundLIME} and Haunschmid et al.~\cite{Haunschmid19TwoLevel} treated spectrograms as images and used rectangular time-frequency segments and superpixels, respectively, as interpretable components. In this work we put forward a musical definition of locality that allows us to create perturbed examples that still sound like music.

\subsection{Musical Definition of Locality}
\label{subsec:musLocality}

One defining property that distinguishes audio processing from image processing is the lack of occlusion in audio. This means that multiple sound sources can be heard simultaneously, whereas overlapping objects in an image cannot all be visualized (the closest visual analogue would be transparency). Another defining feature of audio is that harmonic sources do not produce contiguous regions on a spectrogram, as opposed to visual objects which can be represented contiguously in an image.

We propose a new means of creating interpretable representations other than used previously~\cite{MishraSD17SoundLIME, Haunschmid19TwoLevel}. Instead of applying approaches from image segmentation and treating each ``pixel'' as if it could only belong to one component, we propose using \textit{source separation estimates} as interpretable representations. 

For an auditory mixture defined by a complex valued spectrogram $X \in \mathbb{C}^{F\times T}$, a source separation algorithm is designed to estimate the mix's $C$ constituent sources, $S_c \in \mathbb{C}^{F\times T}$ where all sources $S_c$ can be mixed to make the original mixture $X$. The mixture does not necessarily have to be music: using an appropriate source separation model this approach can also be applied to other domains, e.g. speech~\cite{hershey2016deep} or other acoustic scenes~\cite{kavalerov2019_universal, pishdadian2020learning}, which makes our approach applicable to other audio tasks. 

More specifically, we propose the following method for generating \textit{local} perturbations of a given audio sample: We use the $C$ estimated sources (e.g. corresponding to the piano, the guitar, the vocals, and the bass) of an input audio as our interpretable components. Mapping $z' \in \{0, 1\}^C$ to $z$ (the input audio) is performed by mixing all present sources. In our example $z' = {0, 1, 0, 1}$ results in a music piece only containing the guitar and the bass.
The relation of this approach to the notion of \textit{locality} as used in LIME lies in the fact that samples perturbed in this way will in general still be perceptually similar (i.e., recognized by a human as referring to the same audio piece).
A sketch of the architecture is shown in Figure~\ref{fig:audioLIME}.

In addition to estimating the sources of the input audio, we can also apply \textit{temporal segmentation} into $\tau$ segments, as done in~\cite{MishraSD17SoundLIME}. An example would then be represented by $C \times \tau$ interpretable components. 
Depending on $C$ (between 4 and 6 sources are estimated in our experiments) and $\tau$ (currently set to 1, leaving temporal segmentation for future work) the number of interpretable components can be very small. This leads to a small neighborhood from which to draw the perturbed samples; in our experiments it even allows us to use every example in the neighborhood (between $2^4$ and $2^6$). This actually helps avoiding one of the main issues with LIME -- the instability when repeating the sampling process~\cite{molnar2019} -- as we do not need to sample large numbers of perturbations (as proposed by~\cite{Haunschmid19TwoLevel}, who had to deal with about 300 interpretable components and 50k sampled instances per explanation).

Due to the fact that we are using the same interpretable components for each test example we can draw conclusions about a larger set of examples instead of looking at them individually. This property will be useful for many of our experiments, as it permits us to count, for example, how often a source (e.g. drums) is responsible for making a prediction, and visualize the linear models for the whole test set.
We will take a closer look at this feature in the context of music tagging, in Section \ref{sec:global-analysis}.

\section{Source Separation Models}
\label{sec:sourcesep_models}

\begin{table}
    \begin{tabular}{l|l}
    ~                                     & Model \& Estimated sources                         \\ \hline\hline
    \multirow{3}{*}{\makecell{Validation of \\ audioLIME \\ (Section~\ref{sec:validation})}}& oracle (Slakh2100): bs, dr, gt, pn, \textit{st}   \\
    ~                                     & Cerberus: bs, dr, gt, pn, st, res \\
    ~                                     & One-vs-All (OVA): bs, dr, gt, pn, \textit{res}      \\ \hline
    \multirow{2}{*}{\makecell{Demonstration \\ (Section~\ref{sec:demonstration})}} & oracle (MUSDB18): bs, dr, ot, voc                \\
    ~                                     & spleeter:5stems: bs, dr, ot, pn, voc         \\
    \end{tabular}
    \caption{Source separation models used in our experiments, and the sources they estimate (bs: bass, dr: drums, gt: guitar, pn: piano, st: strings, voc: vocals, ot: other, res: residual). Sources in \textit{italic} are not used in all experiments.}
    \label{tbl:sourcesepmodels}
\end{table}

In our audioLIME experiments, we employ a number of different source separation algorithms, as well as oracle methods (an overview can be found in Table~\ref{tbl:sourcesepmodels}). 

For the experiments in Section~\ref{sec:validation}, we use three types of models. The first is a set of mask inference models trained in the same way as proposed
in~\cite{ManilowWSR19slakh}. 
In this case, one model is trained to estimate a mask for a single source. 
As such we refer to these models as one-vs-all (OVA) models. 

The second type of models used in the Slakh2100 experiments are Cerberus models described by Manilow et al.~\cite{Manilow19Cerberus}. These are multi-source models, meaning that output masks for many sources are estimated at once. 

For some of the experiments using Cerberus and OVA models the ``residual'' (signal left over by source separation) -- computed by subtracting the sum of all estimates from the mix -- was used as an additional component. The idea behind this is to ensure that an interpretable representation with all components present results exactly in the input mix when mapped back to the input space. Our experiments showed, however, that the residual is negligible and does not change the results.

Finally, for the oracle separation ``models'' to be used to obtain baseline results, we use the ground truth sources of the Slakh2100 dataset instead of output from the source separation models.

For the demonstration using real recordings (see Section~\ref{sec:demonstration}), we also use a source separation system trained on real recordings, namely, Spleeter~\cite{spleeter2019}. Spleeter is a set of OVA models, each based on a U-Net architecture for source separation~\cite{jansson2017singing} and trained on a large, proprietary corpus of real recordings. 
For the oracle separation ``models'' in this setting, we use the ground truth sources of the MUSDB18~\cite{rafii18musdb} dataset.


\section{Basic Validation of audioLIME}
\label{sec:validation}

The difficulty in evaluating algorithms for interpreting machine learning models stems from the fact that there is no objective way of measuring the quality of model explanations. Just because an explanation makes sense (or appears convincing) to a human does not necessarily mean that it shows what the model actually learned~\cite{AdebayoGMGHK18SanityChecks}.

In order to circumvent this we designed a dataset in a way that permits us to control what is in the audio and make sure that it does not contain any unintended signal (e.g. added by a recording device). This experiment is inspired by the ``Husky vs Wolf'' experiment in the original LIME paper, involving a ``bad'' classifier. They showed that the classifier had not learned the different characteristics of those animals, but instead learned that wolf images in the training set contained snow in the background~\cite{Ribeiro0G16}. 

In the following sections we design a similar experiment by creating a musical version of this problem and training a ``bad'' classifier i.e., one that focuses on the confounding factor -- the ``snow'' -- rather than on relevant aspects. We expect this classifier to perform poorly when, at test time, the confounding source happens to be present in the other class. 
First, we will show how audioLIME can help detect that a classifier is basing its decisions on a confounding source. An example (including samples of the mixed audio) is shown on our demonstration website (see Section~\ref{sec:local-analysis}). Second, we will perform a sanity check (as proposed by Adebayo et al.~\cite{AdebayoGMGHK18SanityChecks}) 
that shows that our proposed method indeed behaves randomly on untrained models, rather than producing systematic-looking `pseudo-explanations', as many gradient-based methods do.

\subsection{Datasets and Classification Models}

In order to have full control over what instruments are present or absent in our data, we use the recently published Slakh2100~\cite{ManilowWSR19slakh} dataset, and mix the audio ourselves.
Slakh2100 consists of 2100 mixed songs, corresponding MIDI files, with individual instruments separated into MIDI tracks, and each instrument track also individually rendered into audio (called \textit{stems}). All songs contain at least one piano, bass, guitar and drum track\footnote{There are many types of pianos (electric, acoustic), guitars (acoustic, distorted), etc, so the timbre varies widely for any given instrument.}. The majority  of songs (1758) also contain strings. 




For our musical version of the ``Husky vs Wolf'' experiment we will train a
``Guitar vs Piano'' classifier on audio clips re-mixed from
original Slakh songs in a controlled way, where the \textit{drums}
will be used as the confounding source (the ``Snow''). Specifically,
we mix together the following instrument tracks (stems):
the true target instrument (guitar or piano -- possibly several stems,
if the source song contains several pianos/guitars); the confounder
(drums); and optionally additional instruments (bass, strings), to make
the mixes sound more like realistic music.

In the training and validation sets, the drums \textit{always} co-occur
with guitars and \textit{never} co-occur with pianos. For the test sets,
we create two variants: the first, $t_{\text{dr-gtr}}$, follows the same
criteria as  the training and validation sets, \textit{i.e.}, that drums
always occur with guitars and never with pianos. The second, confounding
test set, named $t_{\text{dr-pno}}$, is the opposite: drums always
co-occur with \textit{pianos} and never with guitars.
The class distribution (guitar/piano) is made to be 50:50 in all
data sets.

Finally, we create two versions of these datasets:
the mixes in one version also include a bass track (models trained on
this version will be marked ``gpdb''), the second includes both bass
and strings (``gpdbs'').
That gives us, finally, 1 training, 1 validation, and 2 test sets
for each of the two scenarios (gpdb and gpdbs).

Audio for the training, validation and test sets was pulled from the \textit{Slakh2100-orig} split. The input audio is resampled to 16 kHz and silence below 60 dB is removed. We then compute the STFT (\texttt{n\_fft=1024}, \texttt{hop\_length=512}), log scale the magnitude and compute mel spectrograms, with $128$ frequency bins. For model training we randomly select 2 second snippets (different snippets per training epoch).




\begin{table}[t]
    \centering
    \newcolumntype{C}{ @{}>{${}}c<{{}$}@{} }
    \begin{tabular}{l| *2{rCl}}
    \multicolumn{1}{c|}{\multirow{2}{*}{Model}} & \multicolumn{6}{c}{Test set accuracy ($\% \pm$ SD)} \\ 
      & \multicolumn{3}{c}{{\large $t_{\text{dr-gtr}}$}}  &  \multicolumn{3}{c}{{\large $t_{\text{dr-pno}}$}} \\ \thline
    VGG$_{\text{gpdb}}$ & 
    $99.87\phantom{-}$ & 
    $$\pm$$ & 
    \multicolumn{1}{l|}{$0.28$} & 
    $12.58$\phantom{-} & 
    $$\pm$$ & 
    \phantom{-}$2.98$  \\
    ${\text{VGG}}_{\text{gpdbs}}$ & $99.24$\phantom{-} & $$\pm$$ & \multicolumn{1}{l|}{$0.69$} & $4.58$\phantom{-} &  $$\pm$$ & \phantom{-}$2.71$  \\
    \end{tabular}
        \caption{Accuracy (over 10 runs) of two $\text{VGG16\_bn}$ networks trained to classify mixes with piano vs. mixes with guitar. In training, drums always co-occur with guitar and never piano. This is also true for test set $t_{\text{dr-gtr}}$, while $t_{\text{dr-pno}}$ is the opposite: drums always co-occur with piano.}
    \label{tbl:GPPerformance}
\end{table}

We trained two classifiers using the two datasets described above, named ${\text{VGG}}_{\text{gpdb}}$ and ${\text{VGG}}_{\text{gpdbs}}$ respectively. The architecture used is a VGG16\_bn~\cite{SimonyanZ14aVGG} with only one input channel.
The models were trained using stochastic gradient descent (SGD) with a learning rate of 0.01 and a batch size of 16. These models were set to train for a maximum of 200 epochs, but had an early stopping criterion with patience 20. 


The results on the test sets are summarized in Table~\ref{tbl:GPPerformance}. The performance drop between the two test sets ($t_{\text{dr-gtr}}$ to $t_{\text{dr-pno}}$) is massive. Clearly the models have learned something other than ``Guitar vs Piano''; we can use audioLIME to determine what exactly these models are focusing on.

\subsection{Detecting the Confounding Class}
\label{subsec:confounding}

\begin{table}[t]
    \centering
    \newcolumntype{C}{ @{}>{${}}c<{{}$}@{} }
    \begin{tabular}{l| *2{rCl}}
    \multicolumn{1}{c|}{\multirow{2}{*}{Separation Method}} & \multicolumn{6}{c}{\% where Drums most important} \\ 
    \multicolumn{1}{c|}{}   & \multicolumn{3}{c}{${\text{VGG}}_{\text{gpdb}}$}  & \multicolumn{3}{c}{${\text{VGG}}_{\text{gpdbs}}$}         \\\thline
    Oracle    & $99.90$\phantom{-} & $$\pm$$ & \multicolumn{1}{l|}{$0.29$} & $99.41$\phantom{-} & $$\pm$$ &\phantom{-} $0.90$ \\
    $\text{Cerberus}_{+ res}$~\cite{Manilow19Cerberus} & $99.90$\phantom{-}& $$\pm$$ & \multicolumn{1}{l|}{$0.29$} & $99.90$\phantom{-} & $$\pm$$ &\phantom{-} $0.30$ \\
    OVA~\cite{ManilowWSR19slakh} & $100.00$\phantom{-} & $$\pm$$ & \multicolumn{1}{l|}{$0.00$} & $99.89$\phantom{-} & $$\pm$$ &\phantom{-} $0.31$ \\
    $\text{OVA}_{+ res}$ & $99.61$\phantom{-} & $$\pm$$ & \multicolumn{1}{l|}{$0.65$} & $99.60$\phantom{-} &  $$\pm$$ & \phantom{-} $0.50$ \\
    \end{tabular}
        \caption{Percentage ($\pm$ SD) that audioLIME determined `drums' was the most important component when predicting \textit{guitar} on $t_{\text{dr-pno}}$, for different source separation approaches (mean over 10 runs). 
        ({\scriptsize \textit{+res}} indicates residual source included)
        }
    \label{tbl:GPDrumsDetected}
\end{table}

\begin{figure*}[!htb]
\begin{subfigure}{.49\textwidth}
    \centering
    \includegraphics[trim={1cm 2cm 2.5cm 2.5cm},clip,width=1\textwidth]{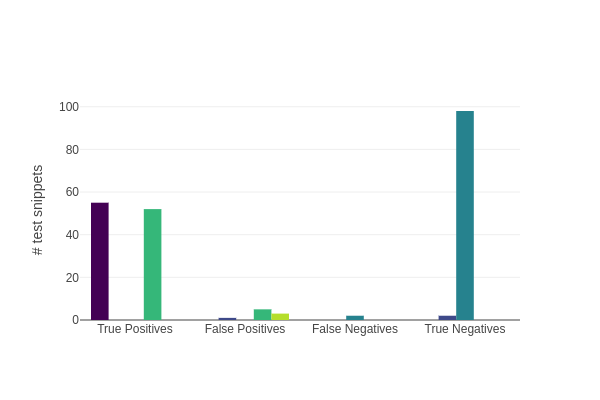}
    \label{fig:important_components_std}
\end{subfigure}
\begin{subfigure}{.49\textwidth}
    \centering
    \includegraphics[trim={1cm 2cm 2.5cm 2.5cm},clip,width=1\textwidth]{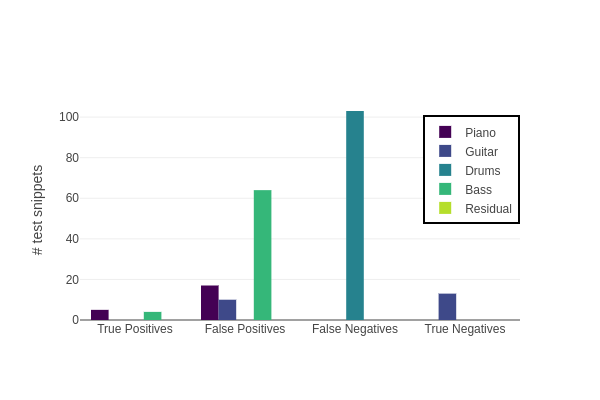}
    \label{fig:important_components_conf}
\end{subfigure}
\caption{Trained ${\text{VGG}}_{\text{gpdb}}$ performance on $t_{\text{dr-gtr}}$ (left) and $t_{\text{dr-pno}}$ (right) and the most important source components as determined by audioLIME for making each of the predictions. In both cases, audioLIME determined that drums is by far the most important source when the model makes predictions. Negative: Guitar, Positive: Piano.
}
\label{fig:slakh_bs_results}
\end{figure*}

In order to gain insights into what our classifier has learned, we apply audioLIME to each example in $t_{\text{dr-gtr}}$ and $t_{\text{dr-pno}}$ using three different separation approaches (see Table~\ref{tbl:sourcesepmodels}). 
We consider the component (estimated source) that is assigned the highest positive weight by the linear explanation model, and count how often each component is considered most strongly responsible for a prediction. We can now check, for a group of examples that we are interested in (e.g., misclassifications), which `feature' seems to be most responsible. If audioLIME works as expected, we would expect to see that the presence of `drums' can explain most misclassifications.

The result for the classifier ${\text{VGG}}_{\text{gpdbs}}$ analysed using audioLIME with Cerberus as a source separation model is visualized in Figure~\ref{fig:slakh_bs_results}. We can observe the following: First, we can see the difference in model performance on $t_{\text{dr-gtr}}$ and $t_{\text{dr-pno}}$. 
The classifier makes few mistakes on the standard test set but has many false positives (guitar classified as piano) and false negatives (piano classified as guitar) on the confused test set. Second, we can see the components that were picked as the most important ones. In our experiments ``guitar'' is the negative class, ``piano'' is the positive class. The colours indicate which component was most often responsible for making a prediction. For both wrong and correct classifications, `drums' was often the responsible component.


A more detailed summary about how often the confounder was detected for 10 classifiers per dataset using different separation approaches can be seen in Table~\ref{tbl:GPDrumsDetected}. The table shows that `drums' was detected to be responsible for the misclassifications almost every time, regardless of the setup of the source separation system. The oracle, Cerberus, and OVA models all operate differently, and yet audioLIME is able to produce similar and correct outcomes regardless of the separation models. This indicates that audioLIME is robust to underlying changes in the exact source separation method. We leave the principled evaluation of source separation performance vs audioLIME detection for future work.




\subsection{Sanity Checks}

\begin{figure}
    \centering
    \includegraphics[trim={1cm 2cm 2.5cm 2.5cm},clip,width=1\columnwidth]{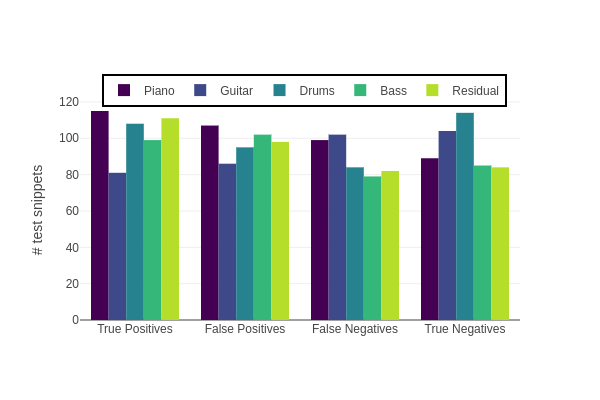}
    \caption{Outcome of the sanity check performed on 10 randomly initialized, untrained ${\text{VGG}}_{\text{gpdb}}$ models, analysing 1 snippet of each test song per model (resulting in 10 predictions and explanations per test song).}
    \label{fig:sanity_check}
\end{figure}

As outlined by Adebayo et al.~\cite{AdebayoGMGHK18SanityChecks}, a simple sanity check for an explanation algorithm is the \textit{model parameter randomization test}. Using this approach, we ask the explanation algorithm under test to create explanations for a randomly initialized, untrained model. For several gradient-based methods the authors showed that the explainer's output often resembles the output of an edge detector, leading to `explanations' that look reasonable to humans, essentially independent of the actual model parameters. If we observe meaningful-looking explanations for an untrained classifier, the explanation method fails the sanity check.

Although the authors state that their sanity checks are also applicable to other explanation methods, for audioLIME-based explanations, looking at single explanations will not reveal much. Applying audioLIME (or LIME in general) to any (trained or untrained) classifier will return some linear weights for each component, which may or may not look reasonable. To draw a conclusion, we need to look at a larger set of examples and visualize the results as in the previous section. If the explanation depends on the model parameters (which it should), we expect to see a difference in the most important components per prediction between an untrained and a trained model. We used 10 randomly initialized ${\text{VGG}}_{\text{gpdbs}}$ models and analysed each of them, in the same way as in Section~\ref{subsec:confounding}, on one randomly selected 2-second snippet per test song (taken from $t_{\text{dr-pno}}$). In contrast to Figure~\ref{fig:slakh_bs_results}, the results in Figure~\ref{fig:sanity_check} show that audioLIME indeed acts randomly as expected on untrained classifiers, proving that it does not create deceptive `explanations'.


\section{Demonstration on Real-world Data: Automated Music Tagging}
\label{sec:demonstration}

\begin{figure*}[!htb]
    \centering
\begin{subfigure}{.585\textwidth}
    \centering
    \includegraphics[width=1\textwidth]{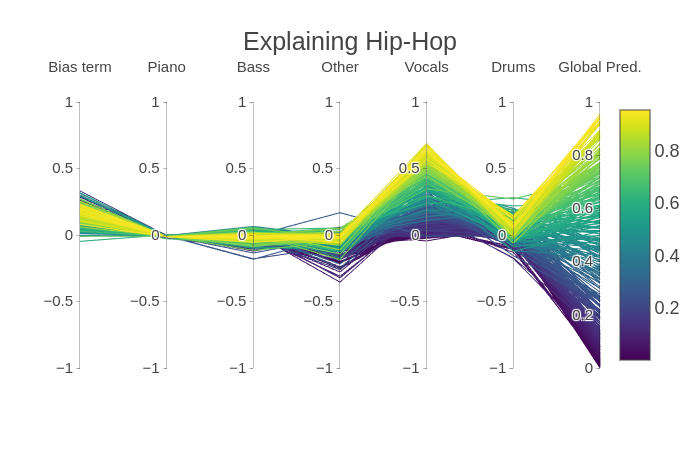}
    \caption{Global analysis of the explanation models for the tag 'Hip-Hop'.}
    \label{fig:paracoords}
\end{subfigure}
\begin{subfigure}{.41\textwidth}
    \centering
    \includegraphics[width=1\textwidth]{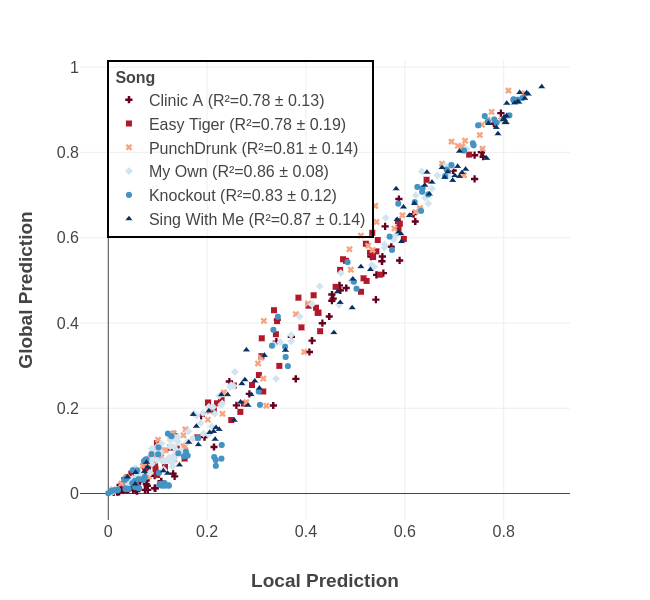}
    \caption{Faithfulness of each linear explanation model.}
    \label{fig:faithfulness}
\end{subfigure}
\caption{This figure shows the results of explaining MSD\_musicnn\_big's predictions of the tag `Hip-Hop' for six songs (478 3-second snippets) from MUSDB18. (a) Each line represents one linear explanation model. Each axis shows the weights for one coefficient, where higher weights indicate that the component was more important for making the prediction. The lines are colored according to MSD\_musicnn\_big's prediction. This nicely shows that `Vocals' contributes most to a high global prediction. (b) This figure shows how well the linear explanation models approximate the global prediction for each snippet. The correlation coefficients summarized over all explanation models per song are shown in the legend.}
\label{fig:hiphop}
\end{figure*}

In a second setting, we analysed the pretrained, automatic music classifier MSD\_musicnn\_big~\cite{pons19musicnn}\footnote{\url{https://github.com/jordipons/musicnn}} which was trained and tested on the Million Song Dataset (MSD)~\cite{Bertin-MahieuxEWL11MSD}). The tagger splits the input song into 3 second snippets and averages the predictions to arrive at predictions for the whole song.
In addition to using the MSD test set for analysing the tagger, we can also ask the tagger to make predictions on other datasets and explain those predictions as well. A suitable dataset for this is MUSDB18~\cite{rafii18musdb}, which contains 150 music tracks with genre annotations and isolated stems. The genre annotations can be compared with the top tag predicted by MSD\_musicnn\_big (out of 50 tags, including genre and instruments, such as \textit{rock}, \textit{guitar}, \textit{female vocalist}, ...) to get an idea of how well the tagger performs on this dataset. As in Section~\ref{sec:validation}, the isolated stems are used as an oracle separation method.

First, we will perform local analysis of a few handpicked examples that seemed interesting. Second, we make use of the unified set of interpretable components, and perform global analysis to ask MSD\_musicnn\_big: What does hip-hop ``sound'' like to you?

\subsection{Local, Case-by-case Analysis}
\label{sec:local-analysis}

For demonstration purposes we show the usage of audioLIME on several different songs\footnote{\label{fn:demo}\url{https://expectopatronum.github.io/demos/towards_musically_meaningful_explanations.html}}. First, we demonstrate audioLIME on two songs of the MUSDB18 dataset and show that using a source separation model in comparison to oracle separation results in picking the same component as most important. Second, on a song that was recorded under real-world conditions (i.e. contains some noise), we show that the top tag `jazz' in distinct parts of the song are all explained by the component `piano'. Though we are careful not to generalize from this local explanation,
it tells us that for this specific song the characteristics of the piano are responsible for making this prediction. From this we can conclude that the model learned to relate certain characteristics of the piano to the tag `jazz'.

\subsection{From Local to Global Model Analysis}
\label{sec:global-analysis}

Instead of only analysing single test examples (local analysis), we can ask our tagger MSD\_musicnn\_big what a certain tag ``sounds'' like to the model. We can do this by running audioLIME on a set of examples (global analysis) and visualizing the weights of all linear explanation models. 

For this global analysis we picked 6 MUSDB18 songs for which the top tag predicted by MSD\_musicnn\_big was `Hip-Hop'. Since it coincided with the genre annotation `rap' in all 6 cases when it was predicted we believe that the model has a relatively good idea what hip-hop/rap ``sounds'' like. We performed audioLIME analysis (using spleeter) on all 3 second snippets of the selected songs\footnote{Note that while the top tag for each of the full songs was `Hip-Hop' this is not the case for each snippet.}. This results in 478 explanations (one per test snippet) whose weights are visualized in a parallel coordinates plot in Figure~\ref{fig:paracoords}. We can observe the importance of `Vocals' for predicting `Hip-Hop', and also that instruments commonly associated with hip-hop like bass and drums are less important by comparison for these examples. 
As in the previous section, it is crucial to point out that this does not imply that the presence of (singing) voice \textit{per se} will result in a high probability of a song being tagged `Hip-Hop'. Regardless, the characteristics of the vocals are strongly related to this genre, which increases our trust that this model has actually learned something about `Hip-Hop' (even if audioLIME's explanation itself cannot tell us exactly which aspect of the voice it is that is important).

In order to rule out introducing artifacts by the source separation model the same analyses were performed using MUSDB18 stems as our oracle separation method and the results look extremely similar. 

To determine how much we can trust those explanations, we can additionally look at the \textit{faithfulness} of our explanation model. Faithfulness is defined as how well the local explanation model approximates the global model under analysis (MSD\_musicnn\_big in this experiment) in the neighborhood of the test example~\cite{Ribeiro0G16}. It can be (a) visualized by comparing the global prediction (by MSD\_musicnn\_big) and the linear explanation model's local prediction and, (b) summarized by the correlation coefficients which LIME automatically computes per linear model. The faithfulness for each snippet of the selected songs is shown in Figure~\ref{fig:faithfulness}.

\section{Conclusion}
\label{sec:conclusion}

We have introduced \textit{audioLIME}, a LIME-based, model-agnostic method that can generate listenable explanations for arbitrary music classification and tagging systems. 
We validated our approach by (a) showing that it can explain the aberrant behaviour of a purposely confused model, and (b) performing (and passing) a sanity check. We demonstrated our method on a popular music tagging model.

All our experiments were performed without temporal segmentation, which is reasonable with the short snippets (2 and 3 seconds, respectively) the models are working with. A fundamental limitation of the explanations audioLIME can produce is the used audio decomposition approach (i.e. source separation), which only allows it to point to the \textit{components} that seem most
important for the  prediction, but not to tell us exactly what
specific \textit{aspects} of these components the classifier
really focuses on.
In future work we will investigate temporal segmentation and other means of audio decomposition to achieve more fine grained explanations and address the question: What is it in each component that is important to the model?

\section{Acknowledgements}

This work has made use of the Mystic (Programmable Systems Research Testbed to Explore a Stack-WIde Adaptive System fabriC) NSF-funded infrastructure at Illinois Institute of Technology, NSF award CRI-1730689.

\bibliography{main}


\end{document}